\documentclass[aps,prl,epsfigure,twocolumn]{revtex4}
\usepackage{amsmath}    
\usepackage{amsfonts}
\usepackage{epsfig}
\usepackage{epstopdf}

\newcommand{\miniket}[1]{\vert#1\rangle}

\newcommand{\minibra}[1]{\langle#1\vert}

\begin{document}

\title{Localization-like effect in two-dimensional alternate quantum walks with periodic coin operations}

\author{Carlo Di Franco$^{1}$ and Mauro Paternostro$^{2}$}

\affiliation{$^{1}$QOLS, Blackett Laboratory, Imperial College London, SW7 2BW, United Kingdom\\$^{2}$Centre for Theoretical Atomic, Molecular and Optical Physics, School of Mathematics and Physics, Queen's University, Belfast, BT7 1NN, United Kingdom}

\begin{abstract}
Exploiting multi-dimensional quantum walks as feasible platforms for quantum computation and quantum simulation is attracting constantly growing attention from a broad experimental physics community. Here, we propose a two-dimensional quantum walk scheme with a single-qubit coin that presents, in the considered regimes, a strong localization-like effect on the walker. The result could provide new possible directions for the implementation of quantum algorithms or from the point of view of quantum simulation. We characterize the localization-like effect in terms of the parameters of a step-dependent qubit operation that acts on the coin space after any standard coin operation, showing that a proper choice can guarantee a non-negligible probability of finding the walker in the origin even for large times. We finally discuss the robustness to imperfections, a qualitative relation with coherences behavior, and possible experimental realizations of this model with the current state-of-the-art settings.
\end{abstract}

\maketitle

The field of quantum computation and quantum simulation has been recently driven to a new rising edge by the experimental realization of quantum walks in various setups, highlighting that different physical systems can be adapted for the implementation of these models. In particular, optical systems have shown their full potential, allowing the experimental demonstration of two-dimensional quantum walks for the first time~\cite{Silberhorn,ourexperiment}, even if several further progresses will surely be obtained also in the other physical scenarios that have been already exploited for the one-dimensional case~\cite{Mechede,experiments}. It is expected that striking results will be experimentally found when the technology will allow proper control of two- or multi-dimensional models.

From the theoretical point of view, the interest in two-dimensional quantum walks has been boosted by the fact that, differently from the one-dimensional version, higher-dimensional schemes ({\it i.e.}, walkers moving on structures with dimension larger than one) can be exploited for the efficient implementation of quantum search algorithms~\cite{GroverSearch}. In particular, the Grover walk has been intensively studied due to its localization feature~\cite{localization}. It has been proved that the non-localized case of the Grover walk can be simulated by a walk (sometimes denoted as alternate quantum walk) where the requirement of a higher dimensionality of the coin space is substituted with the alternance of the directions in which the walker can move~\cite{ourwalk}. Moreover, from the point of view of quantum simulation, increasing the dimension of the lattice on which the walker can move clearly enriches the class of complex systems that can be simulated.

In the quest for more feasible quantum walk models, an important step forward is presented in this paper: a strong localization-like effect ({\it i.e.}, a behavior that reminds, in the investigated time range, the localization as defined in Ref.~\cite{localization}) can be indeed obtained in the modified version of the alternate quantum walk proposed here. This result could pave the way for adapting this scheme to the realization of quantum algorithms or quantum simulation, providing a clear advantage in terms of experimental resources. After a short introduction on the quantum walk studied in this paper, we characterize it in terms of its relevant parameters and show that a proper choice can guarantee a non-negligible probability of finding the walker in the origin even for large times. We consider how imperfections spoil this effect, in order to make our investigation closer to realistic implementations. We then hint at a qualitative relation between the ability of the system to localize and the behavior of coherences established in the state of the particles performing a two-particle equivalent scheme of the walk, which in turns provides information on the way correlations are set up between them. We finally discuss the possible experimental realization of this model with the current state-of-the-art settings in linear optics and cold-atom devices.

Let us consider a quantum system with two degrees of freedom, and thus described by a vector in the composite Hilbert space ${\cal H}={\cal H}_W\otimes{\cal H}_C$. The coin space ${\cal H}_C$ is a two-dimensional Hilbert space spanned by $\{\miniket{0},\miniket{1}\}$ and the walker space ${\cal H}_W$ is an infinite-dimensional Hilbert space spanned by $\{\miniket{x,y}\}$, with $x$ and $y$ assuming all possible integer values. We take as a basis of this space ${\cal H}$ the set $\{\miniket{x,y,c}\}$, with $\miniket{x,y,c}=\miniket{x,y}_W\otimes\miniket{c}_C$; $x$ and $y$ could denote, for instance, the position of a particle ({\it walker}) along the $x$ and $y$ directions, respectively, while $\miniket{c}_C$ is an internal two-level degree of freedom. From now on, we consider $\miniket{0,0}_W$ as the initial state of our walker, and $\miniket{+_y}_C=(\miniket{0}_C+i\miniket{1}_C)/\sqrt{2}$ as the coin one. The last assumption is just for the sake of clarity and is not necessary, as we will point out later on that our results are independent of the initial state of the coin.

The evolution of the system is given by a sequence of conditional shift, coin operations, and phase gates. The effect of the two different conditional shift operations $\hat{S}_x=\sum_{i,j\in \mathbb{Z}}\miniket{i-1,j,0}\minibra{i,j,0}+\sum_{i,j\in \mathbb{Z}}\miniket{i+1,j,1}\minibra{i,j,1}$ and $\hat{S}_y=\sum_{i,j\in \mathbb{Z}}\miniket{i,j-1,0}\minibra{i,j,0}+\sum_{i,j\in \mathbb{Z}}\miniket{i,j+1,1}\minibra{i,j,1}$ is to move the walker on a two-dimensional plane, in a way that depends on the coin state. If we label the position in the $x$ and $y$ directions with increasing numbers from left to right and from bottom to top, respectively, $\hat{S}_x$ moves the walker one step to the left (right) when the coin-component is in the state $\miniket{0}_C$ ($\miniket{1}_C$) and $\hat{S}_y$ moves the walker one step down (up) when the coin-component is in the state $\miniket{0}_C$ ($\miniket{1}_C$). Our coin operation (acting only on the coin space) is the Hadamard gate
$\hat{H}=\frac{1}{\sqrt{2}}
\begin{pmatrix}
1&1\\
1&-1
\end{pmatrix}$,
as in the original quantum walk~\cite{Aharonov:93}. In order to include some dependence on the (discrete) time (similarly to what has been done in Ref.~\cite{banuls}) in the single walk step, we also introduce the phase gate $\hat{P}_{\phi}(t)=
\begin{pmatrix}
e^{-i\frac{\phi}{2}t}&0\\
0&e^{i\frac{\phi}{2}t}
\end{pmatrix}$,
with $t$ the corresponding time step of the quantum walk [also $\hat{P}_{\phi}(t)$ acts only on the coin space]. A single time step consists here of two Hadamard operations, two phase gates and two movements on the $x$ and $y$ directions, according to the sequence: coin operation - phase gate - movement on $x$ - coin operation - phase gate - movement on $y$.

\begin{figure}[t]
\centerline{\bf (a)}
\centerline{\psfig{figure=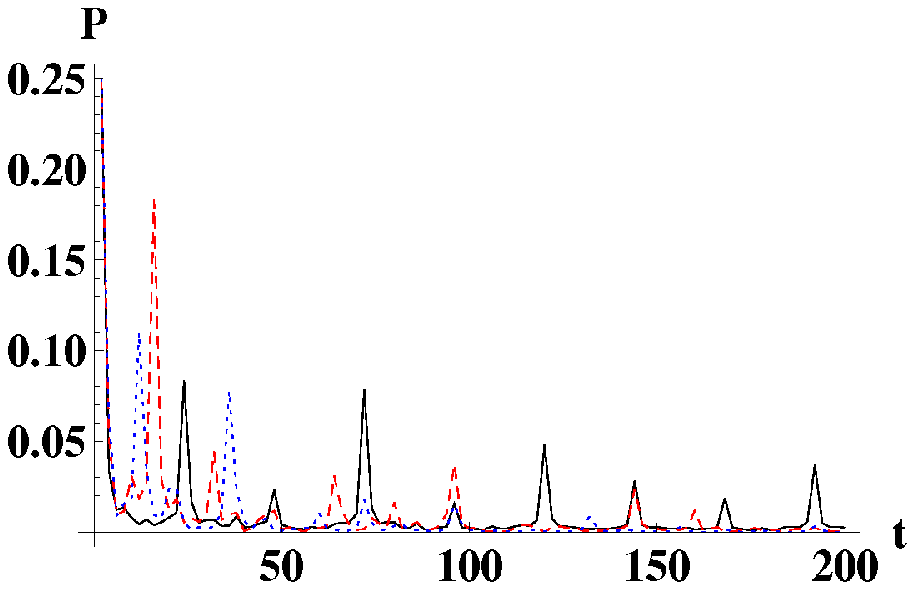,width=6.5cm}}
\vskip1cm
\centerline{\bf (b)}
\centerline{\psfig{figure=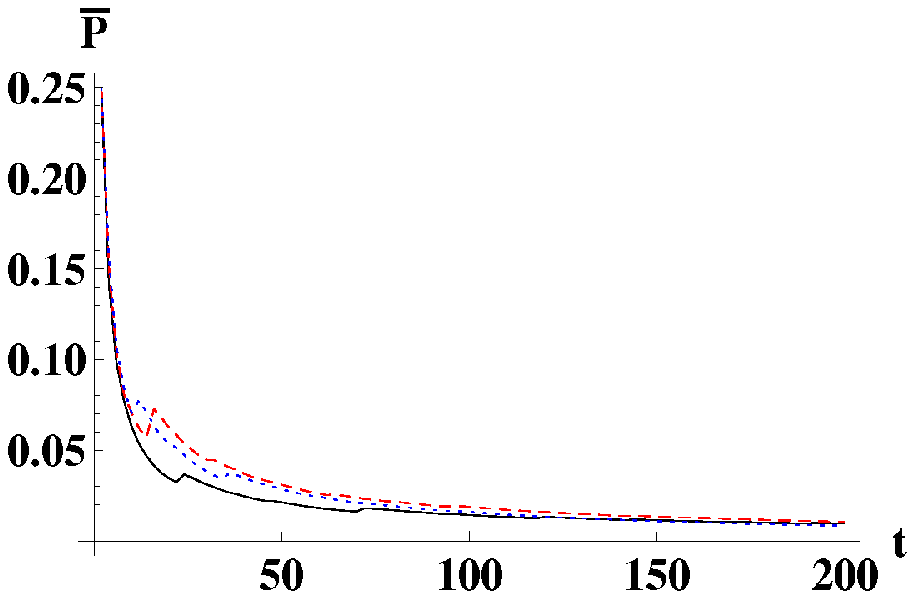,width=6.5cm}}
\caption{{\bf (a)} Probability to find the walker in the origin, against the total number of time steps $t$, for values of $\phi$ equal to $\pi/6$ (black line), $\pi/4$ (red dashed line), $\pi/3$ (blue dotted line). Only the even time steps are shown, as the probability is zero for all odd steps. {\bf (b)} Average probability to find the walker in the origin, against the total number of time steps $t$, for values of $\phi$ equal to $\pi/6$ (black line), $\pi/4$ (red dashed line), $\pi/3$ (blue dotted line).}
\label{plot12}
\end{figure}

Let us first consider the case where both the phase gates (the one before the movement on $x$ and the one before the movement on $y$) have the same $\phi$. We are interested in the spatial probability distribution after a fixed number of steps, that can be obtained by tracing out the state of the coin. In particular, we want to investigate the probability to find the walker in the origin, and check if it quickly decreases with time, or any localization-like effect is present. Fig.~\ref{plot12}{\bf (a)} shows this probability for different values of $\phi$ (we have investigated several other random values of $\phi$, finding always a similar behavior).
For comparison, let us remember that in the case $\phi=0$ ({\it i.e.}, the standard alternate quantum walk) this probability is a non-increasing function of the time. However, for $\phi\ne0$, there are peaks showing a return of the walker in the origin.

It would be interesting to check the walker behavior for long times. In order to do that, we consider the average return probability $\bar{P}$ ({\it i.e.}, the probability to be in the origin after a certain number of steps $t$, averaged over $t$ or, more precisely, over $t/2$, as the probability is zero for all odd steps). We plot it against $t$ in Fig.~\ref{plot12}{\bf (b)}, for the same values of $\phi$ as in Fig.~\ref{plot12}{\bf (a)}. We can notice that the average probability rapidly goes to zero. Also in this case, we have investigated several random values of $\phi$, finding always a similar behavior.

\begin{figure}[t]
\centerline{\bf (a)}
\centerline{\psfig{figure=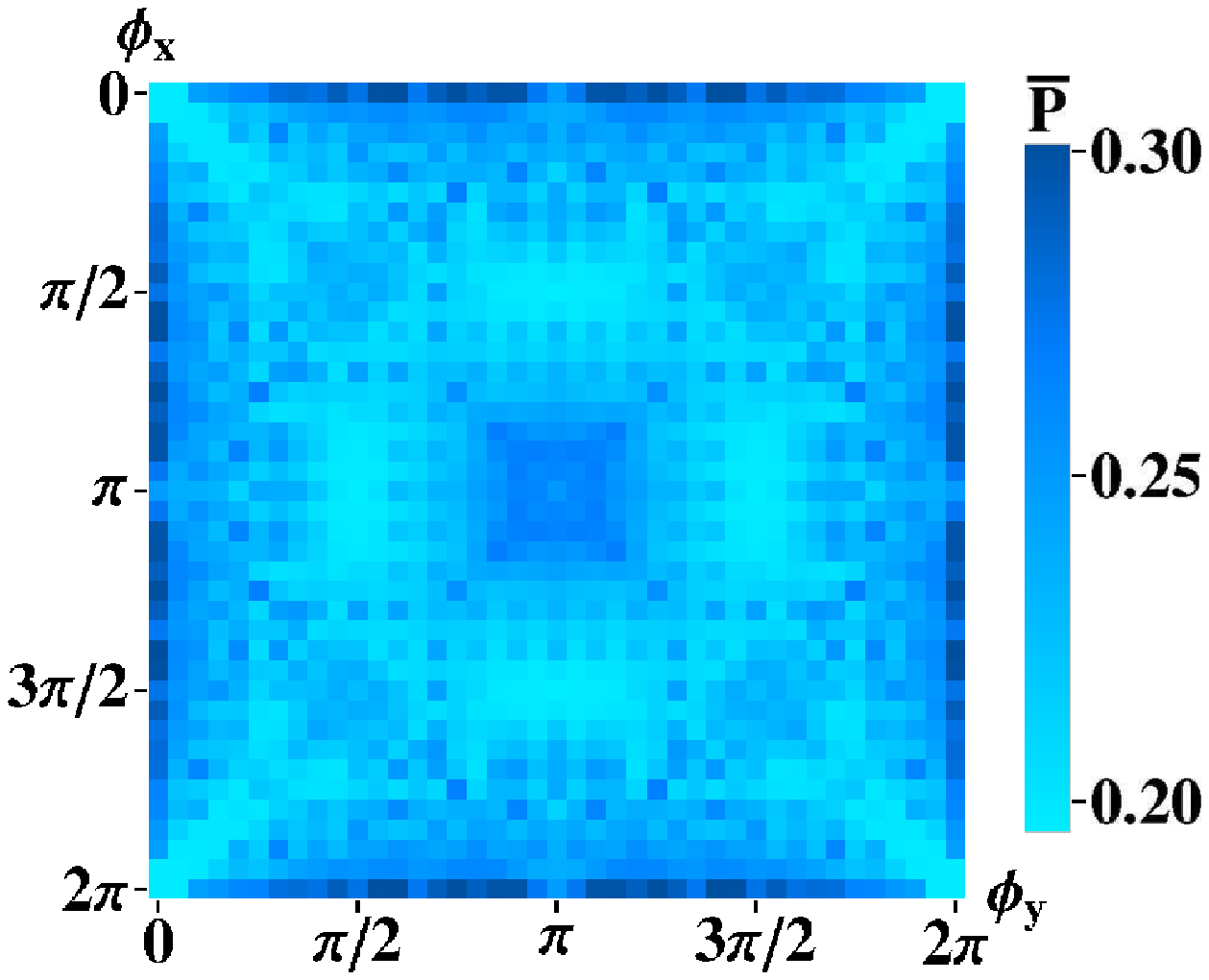,width=6.5cm}}
\vskip1cm
\centerline{\bf (b)}
\centerline{\psfig{figure=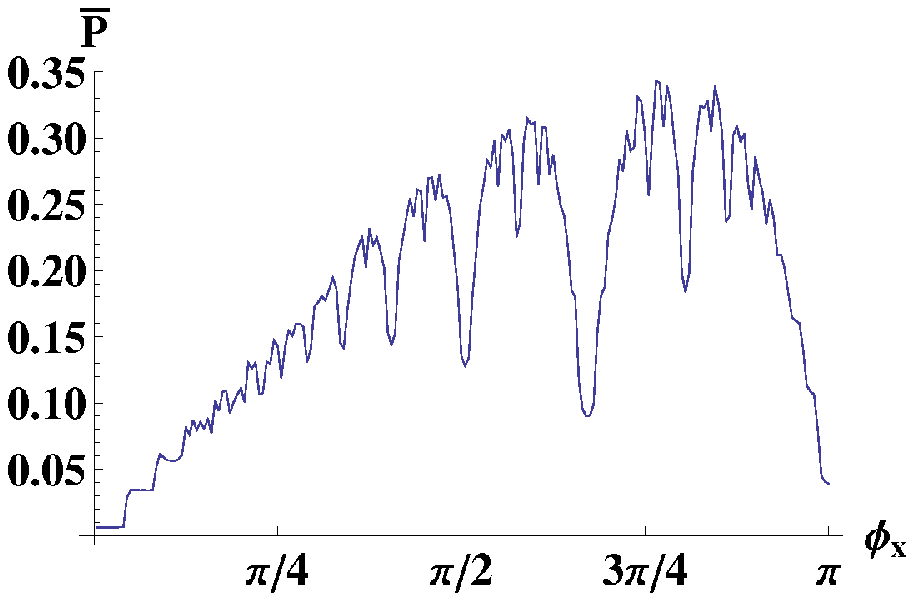,width=6.5cm}}
\caption{{\bf (a)} Average probability to find the walker in the origin, for a total number of time steps $t=40$, against the values of $\phi_x$ and $\phi_y$. {\bf (b)} Average probability to find the walker in the origin, for a total number of time steps $t=100$ and $\phi_y=0$, against the values of $\phi_x$.}
\label{plot34}
\end{figure}

Due to the freedom given by the fact that the scheme investigated here separates the movement on $x$ and $y$, another possibility is clearly to have different phase gates before the two movements, {\it i.e.}, having different $\phi$. Let us define $\phi_x$ ($\phi_y$) the angle in the phase gate before the movement on $x$ (movement on $y$). We can therefore investigate the whole region of the $\phi_x-\phi_y$ plane and check for strong localization-like effects. In order to do that, as a reasonable trade-off between the computational power required by the simulation and the readability of the plot, we fix the total number of steps to $t=40$ and study the average return probability against the two angles, as shown in Fig.~\ref{plot34}{\bf (a)}.
From these results (as well as from other results that we have obtained for different values of the total number of steps $t$) we have found that the maximal values are reached when only one of the two $\phi$'s is different from zero. Moreover, the average probability is symmetric with respect to $\phi_x\leftrightarrow2\pi-\phi_x$, $\phi_y\leftrightarrow2\pi-\phi_y$, and $\phi_x\leftrightarrow\phi_y$.

If we want to find the proper choice of $\phi$ parameters in order to maximize the localization-like effect, and then study its long-time behavior, we can thus just fix one of the $\phi$'s equal to zero (in the following investigation, we choose $\phi_y=0$), and vary the other. This will allow us to reduce the number of parameters and investigate longer times. We plot the obtained results in Fig.~\ref{plot34}{\bf (b)}, where also more points have been taken for the values of $\phi_x$. The total number of steps is $t=100$, in this case. We notice that there are wide regions where the average return probability is much larger than the value corresponding to $\phi_x=\phi_y=0$ ({\it i.e.}, the stardard alternate quantum walk, for which the value is around $0.006$). For comparison, for the same number of time steps, when $\phi_x=\phi_y\ne 0$ (the case analyzed previously), the average return probability never goes above $0.1$. We have already evidence that this choice ($\phi_x\ne 0$, $\phi_y=0$) could guarantee a strong localization-like effect.

\begin{figure}[t]
\centerline{\bf (a)}
\centerline{\psfig{figure=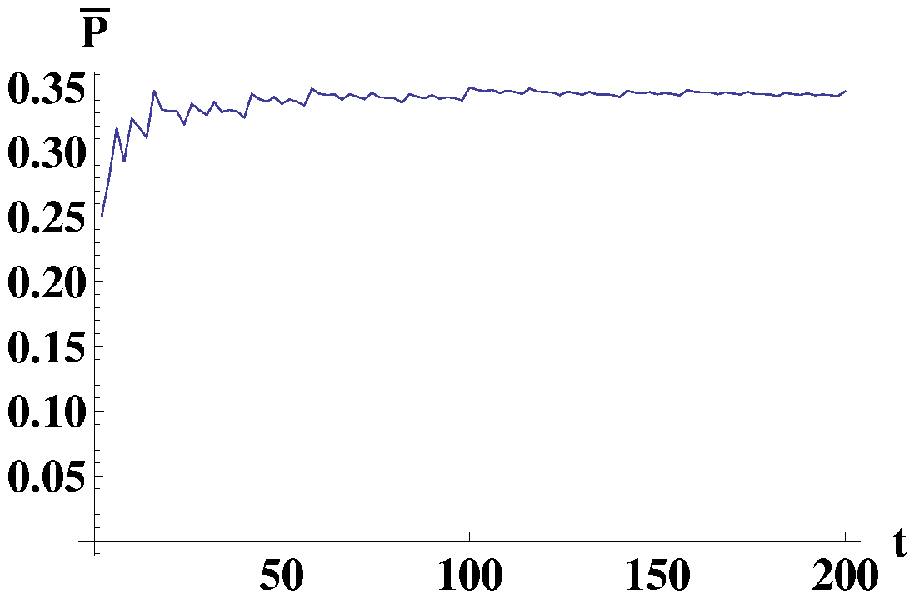,width=6.5cm}}
\vskip1cm
\centerline{\bf (b)}
\centerline{\psfig{figure=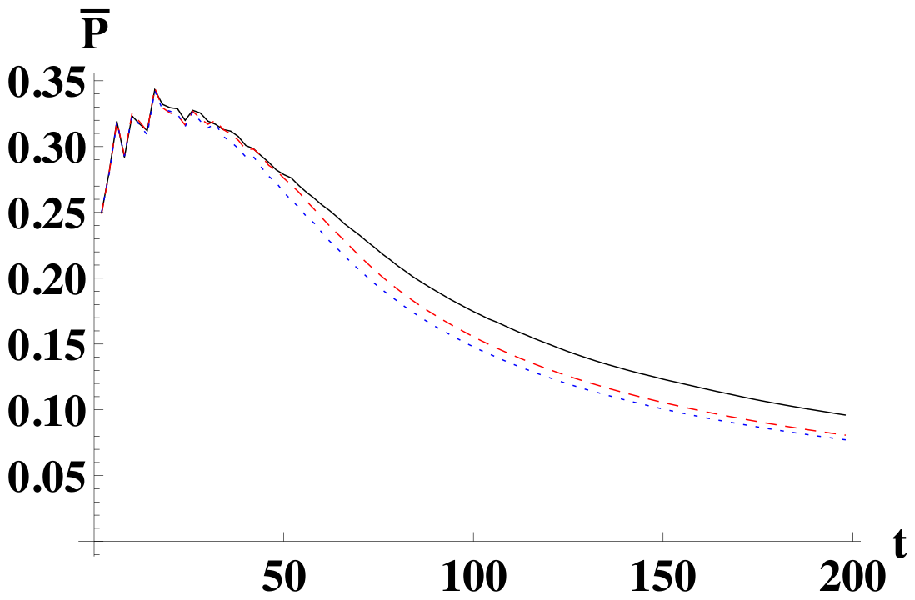,width=6.5cm}}
\caption{{\bf (a)} Average probability to find the walker in the origin, against the total number of time steps $t$, for $\phi_x=19\pi/25$ and $\phi_y=0$. {\bf (b)} Average probability to find the walker in the origin, against the total number of time steps $t$, for a value of $\phi_x=(1+\delta)\phi_0$, $\phi_0=19\pi/25$, $\delta$ sampled at each time step from a uniform distribution in the range $[-10^{-2},10^{-2}]$, and $\phi_y=0$. The black line corresponds to time-dependent disorder, the red dashed line corresponds to position-dependent disorder, the blue dotted line corresponds to disorder depending both on position and time.}
\label{plot56}
\end{figure}

We thus take the value for which we have the maximum at $t=100$ ($\phi_x=19\pi/25$) and we plot, in Fig.~\ref{plot56}{\bf (a)}, the average return probability against $t$ [as done in Fig.~\ref{plot12}{\bf (b)}] for this case. 
The difference is evident: the probability does not go rapidly to zero and remains high for a very long time. It is interesting to notice that a localization-like effect has been observed in one-dimensional quantum walk with periodic coin operation as well~\cite{banuls}. In that case, the effect was just a transition behavior before the walker started to spread. One could speculate that the same happens also in two- or higher-dimensional alternate quantum walks. However, simulating extremely long times for these models is very demanding in terms of computational power. Moreover, the transition time could be several orders of magnitude larger than any time that can be reached in realistic implementations so, for all practical purposes, here we are more interested in long times than in asymptotic limits. We have also extended the analysis for finding the best parameter in the phase gate to the one-dimensional case, as it was not done before. We obtained, even in this case, a behavior similar to that presented in Fig.~\ref{plot34}{\bf (b)}. A localization effect (or also a localization-like effect for a sufficiently long transition period) could strikingly widen the possible applications of the alternate quantum walk. It is worth stressing here that, even if all these results have been calculated by considering the initial state of the coin as $\miniket{+_y}_C=(\miniket{0}_C+i\miniket{1}_C)/\sqrt{2}$, we have checked also the case of other initial coin states. We have found that the average return probability does not depend on the initial state of the coin and therefore our results can be straightforwardly extended to any of them.

If the model has to be used for realistic implementation, it is clearly necessary to analyze how robust it is against imperfections. A systematic error in the value of $\phi_x$ will not strongly affect the localization-like effect, as it can be already noticed in Fig.~\ref{plot34}{\bf (b)}. Even if the value of $\phi_x$ is not exactly the one corresponding to the maximum, wide regions of $\phi_x$ give reasonable return probability. The other possibility is that the error on $\phi_x$ is not systematic. For instance, at each time step the angle could assume a value equal to $(1+\delta)\phi_0$, where $\phi_0$ is the ideal value and $\delta$ is a random variable taking into account possible inaccuracies. In order to check the robustness, we have thus studied the return probability by sampling $\delta$ at each time step from a uniform distribution in the range $[-0.01,0.01]$ (corresponding to a possible error up to $1\%$ of the optimal value). We present the results, averaged over $100$ trails, in Fig.~\ref{plot56}{\bf (b)}.

There are two regions that can be distinguished in the plot. The first one is for short times, where the periodicity in the operation acting on the coin space keeps the return probability quite high, very close to the optimal value in Fig.~\ref{plot56}{\bf (a)}. Then, in the second region, the probability starts to decay, due to the effect of randomization of the phase angle. The value is still higher than the standard alternate quantum walk, but the probability clearly goes to zero for long times. This can be somehow understood by considering that disorder in the standard quantum walk coin operation can have some localization effect as well~\cite{disorderedQW,Silberhorndisorder}. However, three different kinds of disorder can appear: one only depending on the time ({\it i.e.}, a fixed randomization at each time step, similarly to what we have considered previously), one only on the position ({\it i.e.}, a fixed randomization at each position of the lattice on which the walker is moving), and one on both. In particular, only the second gives rise to Anderson localization effects, even if the other two still have localization effects. In all the three cases, however, the probability to find the walker in the origin goes to zero for the time going to infinity. We first want to point out that the case investigated here differs from what studied in the references mentioned above because the disorder in this model is just in the phase gate and not in the general coin operation. However, for the sake of completeness, we have studied the other two cases along the lines of the investigation in Refs.~\cite{disorderedQW,Silberhorndisorder} ({\it i.e.}, $\delta$ only depending on the position, and $\delta$ depending on both time and position) and reported also these results in Fig.~\ref{plot56}{\bf (b)} (red dashed and blue dotted lines, respectively).

It could be interesting to compare the localization phenomenon to the behavior of coherences established between the particles performing a two-particle equivalent scheme of the walk. As described in Ref.~\cite{ourwalkproceeding}, the alternate quantum walk can also been seen from a different point of view: two particles both moving on a line, not interacting directly but sharing a common degree of freedom, embodied by the coin. The one-dimensional movement of the first particle corresponds to the shift of the original walker on the $x$ direction, and the one-dimensional movement of the second particle to the shift of the original walker on the $y$ direction. Intuitively, one expects that the achievement of localization in the alternate walk would affect the way the two walkers are able to correlate each other, establishing some form of localization-radius. The amount of coherences between them could be linked to the dimension of an {\it effective} position Hilbert space. In turn, this would be reflected in the way off-diagonal elements in the two-walker density matrix are populated as the number of steps in our process grows.

\begin{figure}[t]
\centerline{\psfig{figure=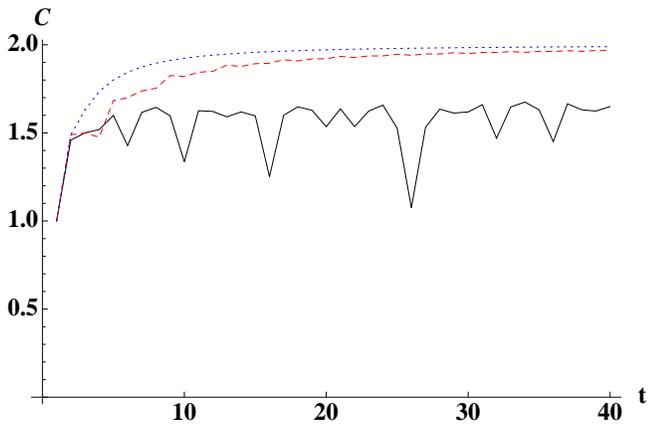,width=8.5cm}}
\caption{Coherence norm against the total number of time steps $t$, for $\phi_x=19\pi/25$, $\phi_y=0$ (black line), $\phi_x=\pi/4$, $\phi_y=0$ (red dashed line), and $\phi_x=0$, $\phi_y=0$ (blue dotted line).}
\label{plot7}
\end{figure}

In order to provide a quantitative assessment of such expectations, we have studied the behavior of the quantity ${\cal C}=\|\rho_W-\rho_{W,x}\otimes\rho_{W,y}\|$ with $\|A\|=\text{Tr}\sqrt{A^\dag A}$ the trace-norm of an operator $A$, $\rho_{W}$ the density matrix of the walkers, and $\rho_{W,k}~(k=x,y)$ its reduction for the walker along the $k$ axis. We dub ${\cal C}$ as the {\it coherence norm}. By providing the sum of the moduli of the off-diagonal elements of the density matrix $\rho_W$, the coherence norm is adopted, in our study, as a quantifier of the global behavior of walker-walker coherences. Notice that a very recent investigation has put forward similar tools as genuine measures of coherence in a quantum system~\cite{baumgratz}. We have compared the trend followed by the coherence norm for three different arrangements of the walk, namely one case of strong localization ($\phi_x=19\pi/25$, $\phi_y=0$, black line), one case of weak localization ($\phi_x=\pi/4$, $\phi_y=0$, red dashed line), and one case of no-localization ($\phi_x=0$, $\phi_y=0$, blue dotted line). The results, presented in Fig.~\ref{plot7}, show that ${\cal C}$ displays rather distinctive features and confirm our predictions (we have investigated several other random values of $\phi_x$ and $\phi_y$, finding always a similar behavior), suggesting in a quantitative and measurable way a strong connection between the phenomenology of inter-walker coherences and the features of the walk itself.

Let us now discuss possible experimental implementations of this model. The standard alternate quantum walk has been recently realized in a optical loop setup~\cite{ourexperiment}. In order to obtain the model proposed here, we should add a phase gate after the first coin operation (denoted as ``coin 1'' in Figure 1 of Ref.~\cite{ourexperiment}). This can be implemented by means of an electro-optic modulator (EOM), as done in Ref.~\cite{Silberhorndisorder} for demonstrating the effects of disorder in one-dimensional case. As shown in Ref.~\cite{Silberhorndisorder}, the EOM can be properly programed in order to change the phases between single steps of the quantum walk, therefore allowing the dependency of the phase gate operation on the time step, as required in the proposed model. Clearly, this time-dependent alternate quantum walk can also be realized in any other physical setups suitable for the standard alternate version. For instance, when an alternate quantum walk will be implemented by means of neutral atoms in optical lattices, along the lines of the one-dimensional experiment in Ref.~\cite{Mechede}, the only change to obtain the model studied here will be in the operation acting on the internal degree of freedom ({\it i.e.}, in the pulse allowing transitions between the considered hyperfine levels of the atoms).

We have studied a time-dependent alternate quantum walk obtaining, in the considered regimes, a strong localization-like effect on the walker. We have investigated this behavior and found the optimal strategy to enhance it. This could pave the way for adapting the scheme to the realization of feasible quantum algorithms, providing a clear advantage in terms of experimental resources. Moreover, it opens new possible directions for enlarging the class of complex systems that can be simulated by this quantum model. We have also shown that the walk presented here can be experimentally implemented with the current state-of-the-art technology and therefore we expect that these results will be experimentally demonstrated in a very near future.

\noindent
{\it Acknowledgments.--}
MP is grateful to S. F. Huelga for invaluable discussions. The authors thank the UK EPSRC (EP/K034480/1 and EP/G004579/1) and the John Templeton Foundation (grant ID 43467).
%
%MP is grateful to S. F. Huelga for invaluable discussions on the issue of coherence in quantum processes. CDF thanks the UK EPSRC (EP/K034480/1). MP acknowledges financial support from the UK EPSRC under the Career Acceleration Fellowship and ``New Directions for EPSRC Research Leaders'' schemes (EP/G004579/1), and the John Templeton Foundation (grant ID 43467).
%

\end{document}